\documentclass[runningheads]{llncs} 

\usepackage{amsmath}
\usepackage{booktabs}
\usepackage{makecell}
\usepackage{tabularx}

\usepackage{todonotes}
\usepackage{xspace}
\usepackage{xcolor,colortbl}
\newcommand{\GDGD}{{$(GD)^2$}\xspace}
\usepackage[colorlinks=true,urlcolor=blue]{hyperref}

\usepackage{float}
\restylefloat{table}

\usepackage{cite}  

\usepackage{caption}
\usepackage{comment}

\newcommand{\reyan}[1]{{\color{red} #1}}

\title{Graph Drawing via Gradient Descent, \GDGD}
\author{Reyan Ahmed\orcidID{0000-0001-6830-9053}, Felice De Luca\orcidID{0000-0001-5937-7636}, Sabin Devkota\orcidID{0000-0002-0610-6573},  Stephen Kobourov\orcidID{0000-0002-0477-2724}, Mingwei Li\orcidID{0000-0002-0457-8091}}
\authorrunning{R. Ahmed, F. De Luca, S. Devkota,  S. Kobourov, M. Li}

\institute{Department of Computer Science, University of Arizona, USA}

\begin{document}


\maketitle
\begin{abstract}
Readability criteria, such as distance or neighborhood preservation, are often used to optimize node-link representations of graphs to enable the comprehension of the underlying data. 
With few exceptions, graph drawing algorithms typically optimize one such criterion, usually at the expense of others. We propose a layout approach, Graph Drawing via Gradient Descent, \GDGD, that can handle multiple readability criteria. 
\GDGD can optimize any criterion that can be described by a smooth function.
If the criterion cannot be captured by a smooth function, a non-smooth function for the criterion is combined with another smooth function, or auto-differentiation tools are used for the optimization.
Our approach is flexible and can be used to optimize several criteria that have already been considered earlier (e.g., obtaining ideal edge lengths, stress, neighborhood preservation) as well as other criteria which have not yet been explicitly optimized in such fashion (e.g., vertex resolution, angular resolution, aspect ratio).
We provide quantitative and qualitative evidence of the effectiveness of \GDGD with experimental data and a functional prototype: \url{http://hdc.cs.arizona.edu/~mwli/graph-drawing/}.

\end{abstract}  

\section{Introduction}
Graphs represent relationships between entities and visualization of this information is relevant in many domains. 
Several criteria have been proposed to evaluate the readability of graph drawings, including the number of edge crossings, distance preservation, and neighborhood preservation. Such criteria evaluate different aspects of the drawing and different layout algorithms optimize different criteria.  It is challenging to optimize multiple readability criteria at once and there are few approaches that can support this.
Examples of approaches that can handle a small number of related criteria include the stress majorization framework of 
Wang et al.~\cite{wang2017revisiting}, which optimizes distance preservation via stress as well as ideal edge length preservation.
The Stress Plus X (SPX) framework of Devkota et al.~\cite{devkota2019stress}  can minimize the number of crossings, or maximize the minimum angle of edge crossings.
While these frameworks can handle a limited set of related criteria, it is not clear how to extend them to arbitrary optimization goals.
The reason for this limitation is that these frameworks are dependent on a particular mathematical formulation. For example, the SPX framework
was designed for crossing minimization, which can be easily modified to handle crossing angle maximization (by adding a cosine factor to the optimization function).
This ``trick" can be applied only to a limited set of criteria but not the majority of other criteria that are incompatible with the basic formulation. 

\begin{figure}[h]
  \includegraphics[width=0.32\textwidth]{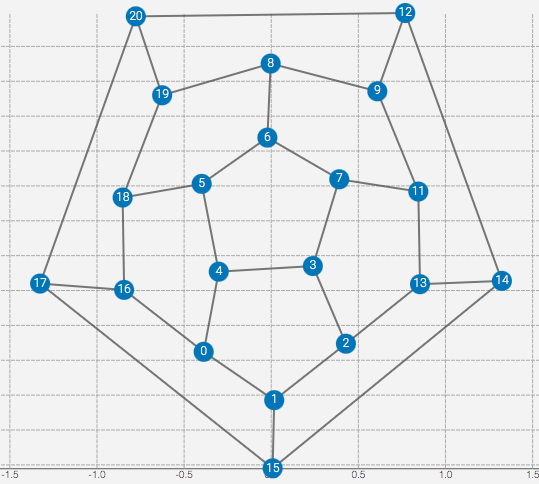}
  \includegraphics[width=0.34\textwidth]{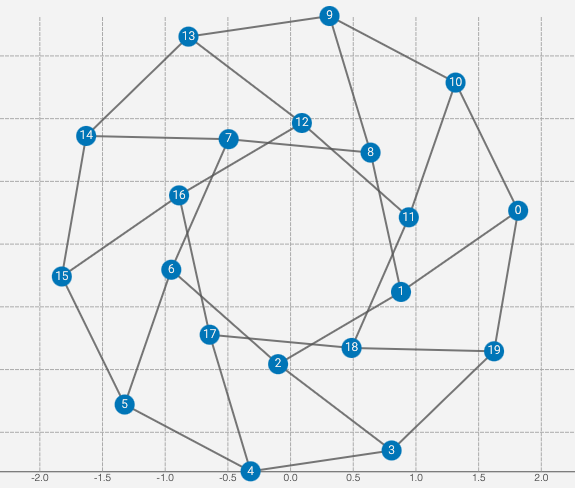}
  \includegraphics[width=0.32\textwidth]{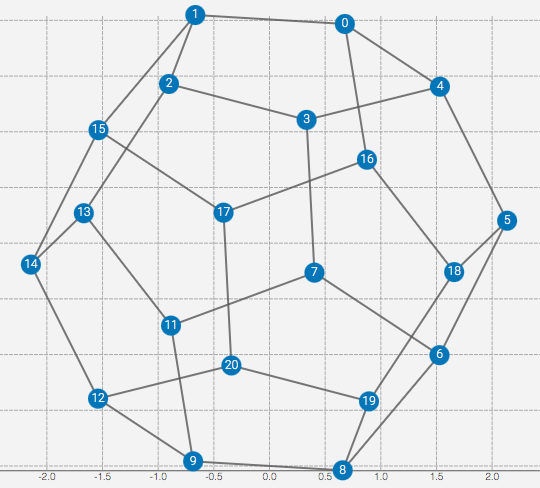}
\caption{Three \GDGD layouts of the  dodecahedron: (a)  optimizing the number of crossings, (b) optimizing uniform edge lengths, and (c) optimizing stress.}
\label{fig:contrast1}
\end{figure}

In this paper, we propose a general approach, Graph Drawing via Gradient Descent, \GDGD, that can optimize a large set of drawing criteria, provided that the corresponding metrics that evaluate the criteria are smooth functions. If the function is not smooth, \GDGD  either combines it with another smooth function and partially optimizes based on the desired criterion, or uses modern auto-differentiation tools to optimize. As a result, the proposed \GDGD framework is simple: it only requires a function that captures a desired drawing criterion.
To demonstrate the flexibility of the approach, we consider an initial set of nine criteria: minimizing stress, maximizing vertex resolution, obtaining ideal edge lengths, maximizing neighborhood preservation, maximizing crossing angle, optimizing total angular resolution, minimizing aspect ratio, optimizing the Gabriel graph property, and minimizing edge crossings. A functional prototype is available on \url{http://hdc.cs.arizona.edu/~mwli/graph-drawing/}. This is an interactive system that allows vertices to be moved manually. Combinations of criteria can be optimized by selecting a weight for each; see Figure~\ref{fig:contrast1}.

\section{Related Work}
Many criteria associated with the readability of graph drawings have been proposed~\cite{ware2002cognitive}. Most of graph layout algorithms are designed to (explicitly or implicitly) optimize a single criterion.
For instance, a classic  layout  criterion is stress minimization ~\cite{kamada_1989}, where stress is defined by $\sum\limits_{i < j}w_{ij} (|X_i-X_j| - d_{ij})^2$. Here, $X$ is a $n\times2$ matrix containing coordinates for the $n$ nodes, $d_{ij}$ is typically the graph-theoretical distance between two nodes $i$ and $j$ and $w_{ij}=d_{ij}^{-\alpha}$ is a normalization factor with $\alpha$ equal to $0, 1$ or $2$. Thus reducing the stress in a layout corresponds to computing node positions so that the actual distance between pairs of nodes is proportional to the graph theoretic distance between them. Optimizing stress can be accomplished by stress minimization, or stress majorization, which can speed up the computation~\cite{gansner2004graph}. In this paper we only consider drawing in the Euclidean plane, however, stress can be also optimized in other spaces such as the torus~\cite{chen2020doughnets}.

Stress minimization corresponds to optimizing the global structure of the layout, as the stress metric takes into account all pairwise distances in the graph. The t-SNET algorithm of Kruiger et al.~\cite{kruiger2017graph} directly optimizes neighborhood preservation, which captures the local structure of a graph, as the 
neighborhood preservation metric only  considers distances between pairs of nodes that are close to each other.
Optimizing local or global distance preservation can be seen as special cases of the more general dimensionality reduction approaches such as  multi-dimensional scaling~\cite{shepard1962analysis,kruskal1964multidimensional}. 

Purchase et al.~\cite{Purchase1997} showed that the readability of graphs increases if a layout has fewer edge crossings. The underlying optimization problem is NP-hard and several graph drawing contests have been organized with the objective of minimizing the number of crossings in the graph drawings~\cite{Abrego12,Buchheim13}. Recently several algorithms that directly minimize crossings have been proposed
~\cite{bennett2010,Radermacher18}. 

The negative impact on graph readability due to edge crossings can be mitigated if crossing pairs of edges have a large crossings angle~\cite{Argyriou2010,huang2014,huang2013,didimo2014crossangle}. Formally, the crossing angle of a straight-line drawing of a graph is the minimum angle between two crossing edges in the layout, and optimizing this property is also NP-hard.
Recent graph drawing contests have been organized with the objective of maximizing the crossings angle in graph drawings and this has led to several heuristics for this problem~\cite{Demel2018AGH,Bekos18}. 


The algorithms above are very effective at optimizing the specific readability criterion they are designed for, but they cannot be directly used to optimize additional criteria. This is a desirable goal, since optimizing one criterion often leads to poor layouts with respect to one or more other criteria: for example, algorithms that optimize the crossing angle tend to create drawings with high stress and no neighborhood preservation~\cite{devkota2019stress}.

Davidson and Harel~\cite{davidson1996drawing} used simulated annealing to optimize different graph readability criteria (keeping nodes away from other nodes and edges, uniform edge lengths, minimizing edge crossings). Recently, several approaches have been proposed to simultaneously improve multiple layout criteria. Wang et al.~\cite{wang2017revisiting} propose a revised formulation of stress that can be used to specify ideal edge direction in addition to ideal edge lengths in a graph drawing. 
Devkota et al.~\cite{devkota2019stress} 
also use a stress-based approach to minimize edge crossings and maximize crossing angles.
Eades et al.~\cite{10.1007/978-3-319-27261-0_41} provided a technique to draw large graphs while optimizing different geometric criteria, including the Gabriel graph property. Although the approaches above are designed to optimize multiple criteria, they cannot be naturally extended to handle other optimization goals. 
 

Constraint-based layout algorithms such as COLA~\cite{ipsepcola_2006, scalable_cola_2009}, can be used to enforce separation constraints on pairs of nodes to
support properties such as customized node ordering or downward
pointing edges. The coordinates of two nodes are related by inequalities in the form of $x_i \geq x_j + gap$ for a node pair $(i,j)$. These kinds of constraints are known as hard constraints and are different from the soft constrains in our \GDGD framework. 

\section{The \GDGD Framework}
%
%

\begin{figure}[t]
\centering
  \includegraphics[width=\linewidth]{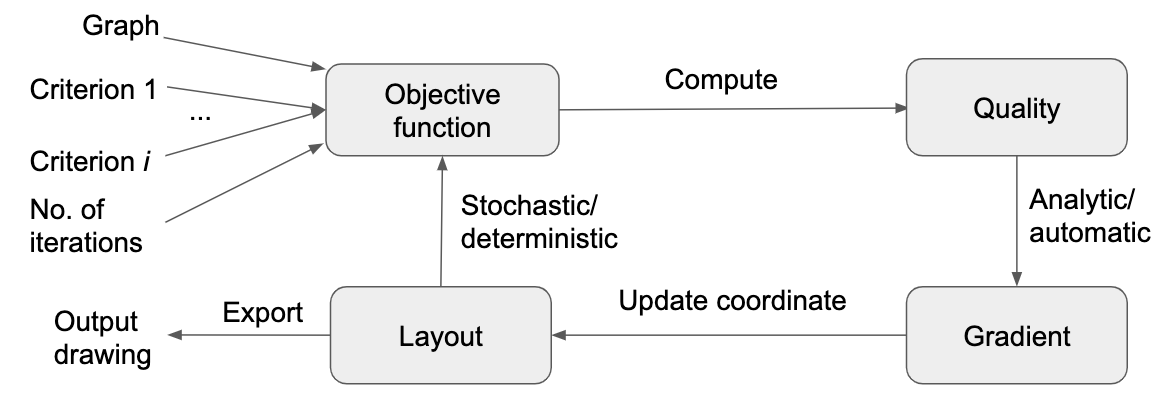}
  \caption{The \GDGD framework: Given a graph and a set of criteria (with weights), formulate an objective function based on the selected set of criteria and weights. Then compute the quality (value) of the objective function of the current layout of the graph. Next, generate the gradient (analytically or automatically). Using the gradient information, update the coordinates of the layout. Finally, update the objective function based on the layout via regular or stochastic gradient descent. This process is repeated for a fixed number of iterations.}
  \label{fig:gdgdframework}
\end{figure}

The \GDGD framework is a general optimization approach to generate a layout with any desired set of aesthetic metrics, provided that they can be expressed by a smooth function. The basic principles underlying this framework are simple. 
The first step is to select a set of layout readability criteria and a loss functions that measures them. 
Then we define the function to optimize as a linear combination of the loss functions for each individual criterion.
Finally, we iterate the gradient descent steps, from which we obtain a slightly better drawing at each iteration.
Figure~\ref{fig:gdgdframework} depicts the framework of \GDGD: 
Given any graph $G$ and readability criterion $Q$, we find a loss function $L_{Q,G}$ which maps from the current layout $X$ (i.e. a $n \times 2$ matrix containing the positions of nodes in the drawing) to a real value that quantifies the current drawing. 
Note that some of the readability criteria naturally correspond to functions that should be minimized (e.g., stress, crossings), while others to functions that should be maximized (e.g., neighborhood preservation, angular resolution). 
Given a loss function $L_{Q,G}$ of $X$ where a lower value is always desirable, at each iteration, a slightly better layout can be found by taking a small ($\epsilon$) step along the (negative) gradient direction: $X^{(new)} = X - \epsilon \cdot \nabla_{X}\; L_{Q,G}$.

To optimize multiple quality measures simultaneously, we take a weighted sum of their loss functions and update the layout by the gradient of the sum.

\subsection{Gradient Descent Optimization}
There are different kinds of gradient descent algorithms. 
The standard method considers all vertices, computes the gradient of the objective function, and updates vertex  coordinates based on the gradient. 
For some objectives, we need to consider all the vertices in every step. 
For example, the basic stress formulation~\cite{kamada_1989} falls in this category.
On the other hand, there are some problems where the objective can be optimized only using a subset of vertices. 
For example, consider stress minimization again. 
If we select a set of vertices randomly and minimize the stress of the induced graph, the stress of the whole graph is also minimized~\cite{zheng2018graph}.
This type of gradient descent is called stochastic gradient descent. 
However, not all objective functions are smooth and we cannot compute the gradient of a non-smooth function. 
In that scenario, we can compute the subgradient, and update the objective based on the subgradient. 
%
Hence, as long as the function is continuously defined on a connected component in the domain, we can apply the subgradient descent algorithm.
In table~\ref{table:loss-functions}, we give a list of loss functions we used to optimize 9 graph drawing properties with gradient descent variants. 
In section~\ref{sect:properties-and-measures}, we specify the loss functions we used in detail.

When a function is not defined in a connected domain, we can introduce a surrogate loss function to `connect the pieces'.
For example, when optimizing neighborhood preservation we maximize the Jaccard similarity between graph neighbors and nearest neighbors in graph layout.
However, Jaccard similarity is only defined between two binary vectors.
To solve this problem we extend Jaccard similarity to all real vectors by its Lov\'{a}sz extension~\cite{berman2018lovasz} and apply that to optimize neighborhood preservation. 
%
An essential part of gradient descent based algorithms is to compute the gradient/subgradient of the objective function. 
In practice, it is always not necessary to write down the gradient analytically as it can be computed automatically via automatic differentiation~\cite{griewank2008evaluating}. 
Deep learning packages such as Tensorflow~\cite{abadi2016tensorflow} and PyTorch~\cite{paszke2019pytorch} apply automatic differentiation to compute the gradient of complicated functions. 
%
%
%

When optimizing multiple criteria simultaneously, we combine them via a weighted sum. 
However, choosing a proper weight for each criterion can be tricky. 
Consider, for example, maximizing crossing angles and minimize stress simultaneously with a fixed pair of weights. 
At the very early stage, the initial drawing may have many crossings and stress minimization often removes most of the early crossings. 
As a result, maximizing crossing angles in the early stages can be harmful as it move nodes in directions that contradict those that come from stress minimization.
Therefore, a well-tailored \textit{weight scheduling} is needed for a successful outcome.
Continuing with the same example, a better outcome can be achieved by first optimizing stress until it converges, and later adding weights for the crossing angle maximization. 
To explore different ways of scheduling, we provide an interface that allows manual tuning of the weights.

\subsection{Implementation}
We implemented the \GDGD framework in JavaScript.
In particular we used the automatic differentiation tools in tensorflow.js~\cite{mlsys2019_154} and the drawing library d3.js~\cite{2011-d3}.
The prototype is available at \url{http://hdc.cs.arizona.edu/~mwli/graph-drawing/}.

\section{Properties and Measures}\label{sect:properties-and-measures}
In this section we specify the aesthetic goals, definitions, quality measures and loss functions for each of the 9 graph drawing properties we optimized: stress, vertex resolution, edge uniformity, neighborhood preservation, crossing angle, aspect ratio, total angular resolution, Gabriel graph property, and crossing number.
In the following discussion, since only one (arbitrary) graph is considered, we omit the subscript $G$ in our definitions of loss function $L_{Q,G}$ and write $L_{Q}$ for short. Other standard graph notation is summarized in Table~\ref{table:notations}.

\begin{table}[]
\begin{tabular}{l|l}
\toprule
Notation & Description\\
\midrule
$G$ & Graph\\
$V$ & The set of nodes in $G$, indexed by $i$, $j$ or $k$\\
$E$ & The set of edges in $G$, indexed by a pair of nodes $(i,j)$ in $V$\\
$n=|V|$ & Number of nodes in $G$\\
$|E|$ & Number of edges in $G$\\
$Adj_{n \times n}$ and $A_{i,j}$ & Adjacency matrix of $G$ and its $(i,j)$-th entry\\
$D_{n \times n}$ and $d_{ij}$ & Graph-theoretic distances between pairs of nodes and the $(i,j)$-th entry\\
$X_{n \times 2}$ & 2D-coordinates of nodes in the drawing\\
$||X_i - X_j||$ & The Euclidean distance between nodes $i$ and $j$ in the drawing\\
$\theta_i$ & $i^{th}$ crossing angle\\
$\varphi_{ijk}$ & Angle between incident edges $(i,j)$ and $(j,k)$\\
\bottomrule
\end{tabular}
\caption{Graph notation used in this paper.}
\label{table:notations}
\end{table}

\subsection{Stress}
We use stress minimization to draw a graph such that the Euclidean distance between pairs of nodes is proportional to their graph theoretic distance. 
Following the ordinary definition of stress~\cite{kamada_1989}, we minimize

\begin{align}
L_{ST} = \sum\limits_{i<j}\;w_{ij}(|X_i - X_j|_2 - d_{ij})^2 \label{eq:loss-stress}
\end{align}
Where $d_{ij}$ is the graph-theoretical distance between nodes $i$ and $j$, $X_i$ and $X_j$ are the 2D coordinates of nodes $i$ and $j$ in the layout. The normalization factor, $w_{ij}=d_{ij}^{-2}$, balances the influence of short and long distances: the longer the graph theoretic distance, the more tolerance we give to the discrepancy between two distances.
When comparing two drawings of the same graph with respect to stress, a smaller value (lower bounded by $0$) corresponds to a better drawing.

\subsection{Ideal Edge Length}
When given a set of ideal edge lengths $\{l_{ij}: (i,j) \in E\}$ we minimize the average deviation from the ideal lengths:

\begin{align}
L_{IL} &= \sqrt{ \frac{1}{|E|} \sum\limits_{(i,j) \in E}\;  
(\frac{||X_i - X_j|| - l_{ij}}{l_{ij}}})^2 \label{eq:loss-ideal-edge-length}
\end{align}
For unweighted graphs, by default we take the average edge length in the current drawing as the ideal edge length for all edges. $l_{ij} = l_{avg} = \frac{1}{|E|} \sum\limits_{(i,j) \in E}\;  ||X_i - X_j||\qquad \text{for all $(i,j) \in E$}$. The quality measure $Q_{IL} = L_{IL}$ is lower bounded by $0$ and a lower score yields a better layout.

\subsection{Neighborhood Preservation}
Neighborhood preservation aims to keep adjacent nodes close to each other in the layout.
Similar to Kruiger et al.~\cite{kruiger2017graph}, the idea is to have the $k$-nearest (Euclidean) neighbors (k-NN) of node $i$ in the drawing to align with the $k$ nearest nodes (in terms of graph distance from $i$). 
A natural quality measure for the alignment is the Jaccard index between the two pieces of information. Let, $Q_{NP} = JaccardIndex(K, Adj) = \frac{|\{(i,j): K_{ij}=1 \text{ and } A_{ij}=1\}|}{|\{(i,j): K_{ij}=1 \text{ or } A_{ij}=1\}|}$, where $Adj$ denotes the adjacency matrix and the $i$-th row in $K$ denotes the $k$-nearest neighborhood information of $i$:
$K_{ij} = 1$ if $j$ is one of the k-nearest neighbors of $i$ and $K_{ij}$ = 0 otherwise.

To express the Jaccard index as a differentiable minimization problem, 
first, we express the neighborhood information in the drawing as a smooth function of node positions $X_i$ and store it in a matrix $\hat{K}$.
In $\hat{K}$, a positive entry $\hat{K}_{i,j}$ means node $j$ is one of the k-nearest neighbors of $i$, otherwise the entry is negative.
Next, we take a differentiable surrogate function of the Jaccard index, the Lov\'{a}sz hinge loss (LHL)~\cite{berman2018lovasz}, to make the Jaccard loss optimizable via gradient descent.
We minimize

\begin{align}
L_{NP} &= LHL(\hat{K}, Adj)\label{eq:lovasz-hinge}
\end{align}
where LHL is given by Berman et al.~\cite{berman2018lovasz}, $\hat{K}$ denotes the $k$-nearest neighbor prediction:

\begin{align}
\hat{K}_{i,j} &= 
\left\{\begin{array}{ll}
  -(||X_i - X_j|| - \frac{d_{i,\pi_k} + d_{i,\pi_{k+1}}}{2} ) & \text{ if } i \neq j\\
  0 & \text{ if } i=j\\
\end{array}\right.\label{eq:neighbor-pred}
\end{align}
where $d_{i,\pi_{k}}$ is the Euclidean distance between node $i$ and its $k^{th}$ nearest neighbor and $Adj$ denotes the adjacency matrix.
Note that $\hat{K}_{i,j}$ is positive if $j$ is a k-NN of $i$, otherwise it is negative, as is required by LHL~\cite{berman2018lovasz}.

\subsection{Crossing Number}
Reducing the number of edge crossings is one of the classic optimization goals in graph drawing, known to affect readability~\cite{Purchase1997}.
Following Shabbeer et al.~\cite{bennett2010}, we employ an expectation-maximization (EM)-like algorithm to minimize the number of crossings. 
Two edges do not cross if and only if there exists a line that separates their extreme points. With this in mind, 
we want to separate every pair of edges (the M step) and use the decision boundaries to guide the movement of nodes in the drawing (the E step).
Formally, given any two edges $e_1=(i,j), e_2=(k,l)$ that do not share any nodes (i.e., $i$, $j$, $k$ and $l$ are all distinct), 
they do not intersect in a drawing (where nodes are drawn at $X_i = (x_i, y_i)$, a row vector) if and only if
there exists a decision boundary $w = w_{(e_1,e_2)}$ (a 2-by-1 column vector) together with a bias $b = b_{(e_1,e_2)}$ (a scalar) such that: $L_{CN, (e_1,e_2)} = \sum_{\alpha = i,j,k \text{ or } l}\;ReLU(1 - t_\alpha \cdot (X_\alpha w + b)) =\; 0$.

Here we use $(e_1,e_2)$ to denote the subgraph of $G$ which only has two edges $e_1$ and $e_2$, $t_i = t_j = 1$ and $t_k = t_l = -1$.
The loss reaches its minimum at $0$ when the SVM classifier $f_{w,b}: x \mapsto xw+b$ predicts node $i$ and $j$ to be greater than $1$ and node $k$ and $l$ to be less than $-1$.
The total loss for the crossing number is therefore the sum over all possible pairs of edges. Similar to (soft) margin SVM, we add a term $|w_{(e_1, e_2)}|^2$ to maximize the margin of the decision boundary: $L_{CN} =\;  \sum\limits_{
\substack{e_1=(i,j),\; e_2=(k,l) \in E\\\text{$i$, $j$, $k$ and $l$ all distinct}}} 
L_{CN, (e_1, e_2)} + |w_{(e_1, e_2)}|^2$. For the E and M steps, we used the same loss function $L_{CN}$ to update the boundaries $w_{(e_1, e_2)}, b_{(e_1, e_2)}$ and node positions $X$:

\begin{align*}
w^{(new)} &= w - \epsilon \nabla_{w} L_{CN} &\text{(M step 1)}\\
b^{(new)} &= b - \epsilon \nabla_{b} L_{CN} &\text{(M step 2)}\\
X^{(new)} &= X - \epsilon \nabla_{X} L_{CN}(X;\; w^{(new)},b^{(new)}) &\text{(E step)}
\end{align*}
To evaluate the quality we simply count the number of crossings.

\subsection{Crossing Angle Maximization}
When edge crossings are unavoidable, the graph drawing can still be easier to read when edges cross at angles close to 90 degrees~\cite{ware2002cognitive}. 
Heuristics such as those by Demel et al.~\cite{Demel2018AGH} and Bekos et al.~\cite{Bekos18} have been proposed and have been successful in graph drawing challenges~\cite{devanny2017graph}.
We use an approach similar to the force-directed algorithm given by Eades et al.~\cite{eades2010force} and minimize the squared cosine of crossing angles: $
L_{CAM} 
= \sum_{\substack{\text{all crossed edge pairs }\\(i,j), (k,l) \in E}} 
(\frac{\langle X_{i}-X_{j}, X_{k}-X_{l}\rangle}{|X_{i}-X_{j}|\cdot|X_{k}-X_{l}|})^2
$. We evaluate quality by measuring the worst (normalized) absolute discrepancy between each crossing angle $\theta$ and the target crossing angle (i.e. 90 degrees):
$
Q_{CAM} = \max_{\theta} |\theta - \frac{\pi}{2}| / \frac{\pi}{2}
$.

\subsection{Aspect Ratio}

Good use of drawing area is often measured by the aspect ratio~\cite{duncan1998balanced} of the bounding box of the drawing, with $1:1$ as the optimum.
We consider multiple rotations of the current drawing and optimize their bounding boxes simultaneously. Let $
AR = \min_{\theta} \frac{\min(w_{\theta}, h_{\theta})}{\max(w_{\theta}, h_{\theta})}
$,
where $w_{\theta}$ and $h_{\theta}$ denote the width and height of the bounding box when the drawing is rotated by $\theta$ degrees.
A naive approach to optimize aspect ratio, which scales the $x$ and $y$ coordinates of the drawing by certain factors, may worsen other criteria we wish to optimize and is therefore not suitable for our purposes.
To make aspect ratio differentiable and compatible with other objectives, we approximate aspect ratio based on $4$ (soft) boundaries (top, bottom, left and right) of the drawing. 
Next, we turn this approximation and the target ($1:1$) into a loss function using  cross entropy loss.
We minimize

\begin{align}
L_{AR} = \sum_{
  \theta \in \{
    \frac{2\pi k}{N}, \text{ for } k=0, \cdots (N-1)
  \}
}\; 
crossEntropy([\frac{w_{\theta}}{w_{\theta}+h_{\theta}}, \frac{h_{\theta}}{w_{\theta}+h_{\theta}}], [0.5, 0.5])\label{eq:aspect-ratio}
\end{align}
where $N$ is the number of rotations sampled (e.g., $N=7$), and $w_{\theta}$, $h_{\theta}$ are the (approximate) width and height of the bounding box when rotating the drawing around its center by an angle $\theta$. For any given $\theta$-rotated drawing, $w_{\theta}$ is defined to be the difference between the current (soft) right and left boundaries, $w_{\theta} = \text{right} - \text{left} = \langle \text{softmax}(x_{\theta}), x_{\theta} \rangle \;-\; \langle \text{softmax}(-x_{\theta}), x_{\theta} \rangle$,
where $x_{\theta}$ is a collection of the $x$ coordinates of all nodes in the $\theta$-rotated drawing, and softmax returns a vector of weights $(\dots w_k, \dots)$ given by 
$
\text{softmax} (x) = (\dots w_k, \dots) = \frac{e^{x_k}}{\sum_i e^{x_i}}
$.
Note that the approximate right boundary is a weighted sum of the $x$ coordinates of all nodes and it is designed to be close to the $x$ coordinate of the right-most node, while keeping other nodes involved.
Optimizing aspect ratio with the softened boundaries will stretch all nodes instead of moving the extreme points.
Similarly, 
$
h_{\theta} = \text{top} - \text{bottom} = \langle \text{softmax}(y_{\theta}), y_{\theta} \rangle \;-\; \langle \text{softmax}(-y_{\theta}), y_{\theta} \rangle
$
Finally, we evaluate the drawing quality by measuring the worst aspect ratio on a finite set of rotations. The quality score ranges from 0 to 1 (where 1 is optimal):
$
Q_{AR} = \min_{
\theta \in \{
    \frac{2\pi k}{N}, \text{ for } k=0, \cdots (N-1)
  \}
} \frac{\min(w_{\theta}, h_{\theta})}{\max(w_{\theta}, h_{\theta})}
$

\subsection{Angular Resolution}
Distributing edges adjacent to a node makes it easier to perceive the information presented in a node-link diagram~\cite{huang2013}. 
Angular resolution~\cite{Argyriou2010}, defined as the minimum angle between incident edges, is one way to quantify this goal. 
Formally, 
$
ANR = \min_{j \in V} \min_{(i,j),(j,k) \in E} \varphi_{ijk}
$,
where $\varphi_{ijk}$ is the angle formed by between edges $(i,j)$ and $(j,k)$.
Note that for any given graph, an upper bound of this quantity is $\frac{2\pi}{d_{max}}$ where $d_{max}$ is the maximum degree of nodes in the graph.
Therefore in the evaluation, we will use this upper bound to normalize our quality measure to $[0,1]$, i.e. 
$
Q_{ANR} = \frac{ANR}{2\pi / d_{max}}
$.
To achieve a better drawing quality via gradient descent, we define the angular energy of an angle $\varphi$ to be $e^{-s \cdot \varphi}$, where $s$ is a constant controlling the sensitivity of angular energy with respect to the angle (by default $s=1$), and minimize the total angular energy over all incident edges:

\begin{align}
L_{ANR} = \sum_{(i,j),(j,k) \in E} e^{-s \cdot \varphi_{ijk}} \label{eq:total-angular-resolution-loss}
\end{align}

\subsection{Vertex Resolution}
Good vertex resolution is associated with the ability to  distinguish different vertices by preventing nodes from occluding each other. 
Vertex resolution is typically defined as the minimum Euclidean distance between two vertices in the drawing~\cite{chrobak1996convex,schulz2011drawing}. 
However, in order to align with the units in other objectives such as stress, we normalize the minimum Euclidean distance with respect to a reference value.
Hence we define the vertex resolution to be the ratio between the shortest and longest  distances between pairs of nodes in the drawing, 
$
VR = \frac{\min_{i \neq j}||X_i - X_j||}{d_{max}}
$,
where $d_{max} = \max_{k,l}||X_k - X_l||$. 
To achieve a certain target resolution $r \in [0,1]$ by minimizing a loss function, we minimize 

\begin{align}
L_{VR} = \sum_{i,j \in V, i \neq j}{ReLU( 1 - \frac{||X_i - X_j||}{r \cdot d_{max}}) \; ^2} \label{eq:loss-vertex-resolution}
\end{align}
In practice, we set the target resolution to be $r=\frac{1}{\sqrt{|V|}}$, where $|V|$ is the number of vertices in the graph. 
In this way, an optimal drawing will distribute nodes uniformly in the drawing area.
The purpose of the ReLU is to output zero when the argument is negative, as when the argument is negative the constraint is already satisfied.
In the evaluation, we report, as a quality measure, the ratio between the actual and target resolution and cap its value between $0$ (worst) and $1$ (best).

\begin{align}
Q_{VR} = \min(1.0, \frac{\min_{i,j} ||X_i - X_j||}{r \cdot d_{max}}) \label{eq:quality-vertex-resolution}
\end{align}

\subsection{Gabriel Graph Property}
A graph is a Gabriel graph if it can be drawn in such a way that any disk formed by using an edge in the graph as its diameter contains no other nodes. Not all graphs are Gabriel graphs, but drawing a graph so that as many of these edge-based disks are empty of other nodes has been associated with good readability~\cite{10.1007/978-3-319-27261-0_41}.
This property can be enforced by a repulsive force around the midpoints of edges.
Formally, we establish a repulsive field with radius $r_{ij}$ equal to half of the edge length, around the midpoint $c_{ij}$ of each edge $(i,j) \in E$, and we minimize the total potential energy:

\begin{align}
L_{GA} = \sum_{
\substack{
  (i,j) \in E,\\
  k \in V \setminus \{i,j\}
}}
ReLU(r_{ij} - |X_k - c_{ij}|) \; ^ 2\label{eq:gabriel}
\end{align}
where 
$
c_{ij} = \frac{X_i + X_j}{2}
$ and
$
r_{ij} = \frac{|X_i - X_j|}{2}
$.
We use the (normalized) minimum distance from nodes to centers to characterize the quality of a drawing with respect to Gabriel graph property:
$
Q_{GA} = \min_{(i,j) \in E, k \in V}\frac{|X_k - c_{ij}|}{r_{ij}}
$.

%

\section{Experimental Evaluation}

\begin{figure}[htbp]
\centering
\includegraphics[trim={25 350 25 60}, clip, width=1.0\linewidth]{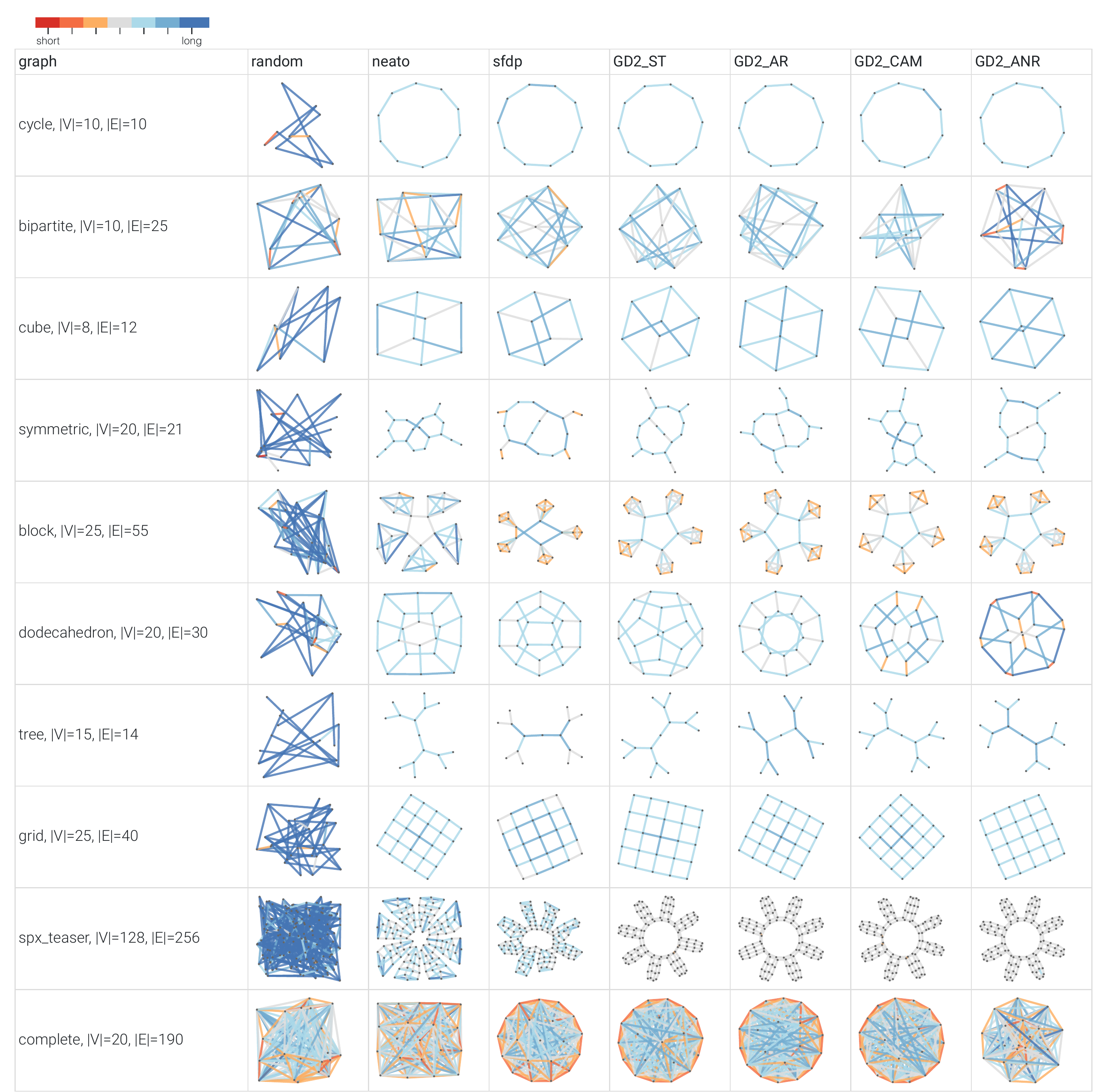}
\caption{Drawings from different algorithms: neato, sfdp and \GDGD with stress (\texttt{ST}), aspect ratio (\texttt{AR}), crossing angle maximization (\texttt{CAM}) and angular resolution (\texttt{ANR}) optimization on a set of 10 graphs. Edge color is determined by the discrepancy between actual and ideal edge length (here all ideal edge lengths are $1$); informally, short edges are red and long edges are blue.}
\label{fig:layoutstable}
\end{figure}

In this section, we describe the experiment we conducted on 10 graphs to assess the effectiveness and limitations of our approach. The graphs used are depicted in Figure~\ref{fig:layoutstable} along with information about each graph. The graphs have been chosen to represent a variety of graph classes such as trees, cycles, grids, bipartite graphs, cubic graphs, and symmetric graphs.



In our experiment we compare \GDGD with neato~\cite{ellson2001graphviz} and  sfdp~\cite{ellson2001graphviz}, which are classical implementations of a stress-minimization layout and scalable force-directed layout. In particular, we focus on 9 readability criteria: stress (\texttt{ST}), vertex resolution (\texttt{VR}), ideal edge lengths (\texttt{IL}), neighbor preservation (\texttt{NP}), crossing angle (\texttt{CA}), angular resolution (\texttt{ANR}), aspect ratio (\texttt{AR}), Gabriel graph properties (\texttt{GG}), and crossings (\texttt{CR}).  We provide the values of the nine criteria corresponding to the 10 graphs for the layouts computed by 
by neato, sfdp, random, and 3 runs of \GDGD initialized with neato, sfdp, and random layouts in Table~\ref{tab:avg_crossing_angle}. 
The best result is shown with bold font, green cells indicate improvement, yellow cells represent ties,
 with respect to the initial values (scores for different criteria obtained using neato, sfdp, and random initialization).
From the experimental results we see that \GDGD improves the random layout in 90\% of the tests.  \GDGD also improves or ties initial layouts from  neato and sfdp, but the improvements are not as strong or as frequent, most notably for the  \texttt{CR}, \texttt{NP}, and \texttt{CA} criteria.

In this experiment, we focused on optimizing a single metric. In some applications, it is desirable to optimize multiple criteria. We can use a similar technique i.e., take a weighted sum of the metrics and optimize the sum of scores. In the prototype (\url{http://hdc.cs.arizona.edu/~mwli/graph-drawing/}), there is a slider for each criterion, making it possible to combine different  criteria.

\section{Limitations}
Although \GDGD is a flexible framework that can optimize a wide range of criteria, it cannot handle the class of constraints where the node coordinates are related by some inequalities, i.e., the framework does not support hard constraints. Similarly, this framework does not naturally support shape-based drawing constraints such as those in~\cite{ipsepcola_2006, scalable_cola_2009,wang2017revisiting}. 
\GDGD takes under a minute for the small graphs considered in this paper. We have not experimented with larger graphs as the implementation has not been optimized for speed.

\section{Conclusions and Future Work}
We introduced the graph drawing framework \GDGD and
showed how this approach can be used to optimize different graph drawing criteria and combinations thereof. The framework is flexible and natural directions for future work include adding further drawing criteria and better ways to combine them.
To  compute the layout of large graphs, a multi-level algorithmic model might be needed. It would also be useful to have a way to compute appropriate weights for the different criteria.

\section*{Acknowledgments}
This work was supported in part by NSF grants CCF-1740858, CCF-1712119, and
DMS-1839274.

\begin{table}[H]
\begin{minipage}{.5\linewidth}
\centering
{\tiny
\begin{tabular}{|l|l|l|l|l|l|l|} \hline
\multicolumn{7}{|c|}{
Crossings
}\\ \hline
\multicolumn{1}{|c|}{} & \multicolumn{1}{c|}{neato} & \multicolumn{1}{c|}{sdfp} & rnd & \multicolumn{1}{c|}{$(GD)^2_n$} & \multicolumn{1}{c|}{$(GD)^2_s$} & \multicolumn{1}{c|}{$(GD)^2_r$} \\ \hline
dodec. 
& \textbf{6.0} 
& \textbf{6.0} 
& 79.0 
& \cellcolor{yellow!50} \textbf{6.0} 
& \cellcolor{yellow!50} \textbf{6.0} 
& \cellcolor{green!50} 10.0 
 \\ \hline
cycle 
& \textbf{0.0} 
& \textbf{0.0} 
& 11.0 
& \cellcolor{yellow!50} \textbf{0.0} 
& \cellcolor{yellow!50} \textbf{0.0} 
& \cellcolor{green!50} \textbf{0.0} 
 \\ \hline
tree 
& \textbf{0.0} 
& \textbf{0.0} 
& 31.0 
& \cellcolor{yellow!50} \textbf{0.0} 
& \cellcolor{yellow!50} \textbf{0.0} 
& \cellcolor{green!50} \textbf{0.0} 
 \\ \hline
block 
& 23.0 
& \textbf{16.0} 
& 297.0 
& \cellcolor{yellow!50} 23.0 
& \cellcolor{yellow!50} \textbf{16.0} 
& \cellcolor{green!50} 25.0 
 \\ \hline
compl. 
& \textbf{3454} 
& 3571 
& 3572 
& \cellcolor{yellow!50} \textbf{3454} 
& \cellcolor{yellow!50} 3571 
& \cellcolor{yellow!50} 3572 
 \\ \hline
cube 
& \textbf{2.0} 
& \textbf{2.0} 
& 18.0 
& \cellcolor{yellow!50} \textbf{2.0} 
& \cellcolor{yellow!50} \textbf{2.0} 
& \cellcolor{green!50} \textbf{2.0} 
 \\ \hline
symme. 
& 1.0 
& \textbf{0.0} 
& 77.0 
& \cellcolor{yellow!50} 1.0 
& \cellcolor{yellow!50} \textbf{0.0} 
& \cellcolor{green!50} \textbf{0.0} 
 \\ \hline
bipar. 
& \textbf{40.0} 
& 52.0 
& \textbf{40.0} 
& \cellcolor{yellow!50} \textbf{40.0} 
& \cellcolor{green!50} \textbf{40.0} 
& \cellcolor{yellow!50} \textbf{40.0} 
 \\ \hline
grid 
& \textbf{0.0} 
& \textbf{0.0} 
& 190.0 
& \cellcolor{yellow!50} \textbf{0.0} 
& \cellcolor{yellow!50} \textbf{0.0} 
& \cellcolor{green!50} \textbf{0.0} 
 \\ \hline
spx t. 
& 73.0 
& \textbf{71.0} 
& 7254.0 
& \cellcolor{yellow!50} 73.0 
& \cellcolor{yellow!50} \textbf{71.0} 
& \cellcolor{green!50} 76.0 
 \\ \hline
\end{tabular}
}
\end{minipage}
\begin{minipage}{.5\linewidth}
\centering
{\tiny
\begin{tabular}{|l|l|l|l|l|l|l|} \hline
\multicolumn{7}{|c|}{
Ideal edge length
}\\ \hline
\multicolumn{1}{|c|}{} & \multicolumn{1}{c|}{neato} & \multicolumn{1}{c|}{sdfp} & rnd & \multicolumn{1}{c|}{$(GD)^2_n$} & \multicolumn{1}{c|}{$(GD)^2_s$} & \multicolumn{1}{c|}{$(GD)^2_r$} \\ \hline
dodec. 
& 0.14 
& 0.15 
& 0.53 
& \cellcolor{green!50} 0.1 
& \cellcolor{yellow!50} 0.15 
& \cellcolor{green!50} \textbf{0.08} 
 \\ \hline
cycle 
& \textbf{0.0} 
& \textbf{0.0} 
& 0.42 
& \cellcolor{yellow!50} \textbf{0.0} 
& \cellcolor{yellow!50} \textbf{0.0} 
& \cellcolor{green!50} \textbf{0.0} 
 \\ \hline
tree 
& \textbf{0.03} 
& 0.13 
& 0.31 
& \cellcolor{yellow!50} \textbf{0.03} 
& \cellcolor{green!50} 0.04 
& \cellcolor{green!50} 0.09 
 \\ \hline
block 
& 0.31 
& 0.43 
& 0.5 
& \cellcolor{green!50} \textbf{0.25} 
& \cellcolor{green!50} 0.33 
& \cellcolor{green!50} 0.31 
 \\ \hline
compl. 
& 0.42 
& \textbf{0.41} 
& 0.45 
& \cellcolor{green!50} \textbf{0.41} 
& \cellcolor{yellow!50} \textbf{0.41} 
& \cellcolor{green!50} \textbf{0.41} 
 \\ \hline
cube 
& 0.08 
& 0.12 
& 0.29 
& \cellcolor{green!50} 0.03 
& \cellcolor{green!50} \textbf{0.0} 
& \cellcolor{green!50} 0.12 
 \\ \hline
symme. 
& 0.08 
& 0.19 
& 0.46 
& \cellcolor{green!50} 0.07 
& \cellcolor{green!50} 0.05 
& \cellcolor{green!50} \textbf{0.04} 
 \\ \hline
bipar. 
& 0.31 
& 0.26 
& 0.44 
& \cellcolor{green!50} 0.16 
& \cellcolor{green!50} 0.13 
& \cellcolor{green!50} \textbf{0.1} 
 \\ \hline
grid 
& 0.01 
& 0.09 
& 0.41 
& \cellcolor{green!50} \textbf{0.0} 
& \cellcolor{green!50} \textbf{0.0} 
& \cellcolor{green!50} 0.01 
 \\ \hline
spx t. 
& 0.4 
& 0.32 
& 0.45 
& \cellcolor{green!50} 0.3 
& \cellcolor{green!50} \textbf{0.2} 
& \cellcolor{green!50} 0.32 
 \\ \hline
\end{tabular}
}
\end{minipage}
\label{tab:avg_edge_uniformity}
\par\medskip
\begin{minipage}{.5\linewidth}
\centering
{\tiny
\begin{tabular}{|l|l|l|l|l|l|l|} \hline
\multicolumn{7}{|c|}{
Stress
}\\ \hline
\multicolumn{1}{|c|}{} & \multicolumn{1}{c|}{neato} & \multicolumn{1}{c|}{sdfp} & rnd & \multicolumn{1}{c|}{$(GD)^2_n$} & \multicolumn{1}{c|}{$(GD)^2_s$} & \multicolumn{1}{c|}{$(GD)^2_r$} \\ \hline
dodec. 
& 21.4 
& 17.58 
& 111.05 
& \cellcolor{green!50} \textbf{17.45} 
& \cellcolor{yellow!50} 17.58 
& \cellcolor{green!50} 17.6 
 \\ \hline
cycle 
& \textbf{0.77} 
& \textbf{0.77} 
& 30.24 
& \cellcolor{yellow!50} \textbf{0.77} 
& \cellcolor{yellow!50} \textbf{0.77} 
& \cellcolor{green!50} \textbf{0.77} 
 \\ \hline
tree 
& \textbf{2.11} 
& 2.7 
& 98.49 
& \cellcolor{yellow!50} \textbf{2.11} 
& \cellcolor{green!50} 2.62 
& \cellcolor{green!50} 5.5 
 \\ \hline
block 
& 26.79 
& 28.22 
& 203.31 
& \cellcolor{green!50} 12.72 
& \cellcolor{green!50} 23.71 
& \cellcolor{green!50} \textbf{11.2} 
 \\ \hline
compl. 
& 33.54 
& 31.58 
& 37.87 
& \cellcolor{green!50} 31.53 
& \cellcolor{green!50} 31.49 
& \cellcolor{green!50} \textbf{31.47} 
 \\ \hline
cube 
& 2.75 
& 2.71 
& 11.69 
& \cellcolor{green!50} 2.66 
& \cellcolor{green!50} 2.69 
& \cellcolor{green!50} \textbf{2.65} 
 \\ \hline
symme. 
& 9.88 
& 5.38 
& 180.48 
& \cellcolor{yellow!50} 9.88 
& \cellcolor{green!50} \textbf{3.36} 
& \cellcolor{green!50} 3.97 
 \\ \hline
bipar. 
& 9.25 
& \textbf{8.5} 
& 12.48 
& \cellcolor{green!50} 8.52 
& \cellcolor{yellow!50} \textbf{8.5} 
& \cellcolor{green!50} 9.6 
 \\ \hline
grid 
& \textbf{6.77} 
& 7.38 
& 221.66 
& \cellcolor{yellow!50} \textbf{6.77} 
& \cellcolor{green!50} 6.78 
& \cellcolor{green!50} \textbf{6.77} 
 \\ \hline
spx t. 
& 674.8 
& 418.4 
& 9794 
& \cellcolor{green!50} \textbf{227.1} 
& \cellcolor{green!50} 235.3 
& \cellcolor{green!50} 227.2
 \\ \hline
\end{tabular}
}
\end{minipage}
\label{tab:avg_stress}
\begin{minipage}{.5\linewidth}
\centering
{\tiny
\begin{tabular}{|l|l|l|l|l|l|l|} \hline
\multicolumn{7}{|c|}{
Angular resolution
}\\ \hline
\multicolumn{1}{|c|}{} & \multicolumn{1}{c|}{neato} & \multicolumn{1}{c|}{sdfp} & rnd & \multicolumn{1}{c|}{$(GD)^2_n$} & \multicolumn{1}{c|}{$(GD)^2_s$} & \multicolumn{1}{c|}{$(GD)^2_r$} \\ \hline
dodec. 
& 0.39 
& 0.39 
& 0.01 
& \cellcolor{green!50} \textbf{0.6} 
& \cellcolor{yellow!50} 0.39 
& \cellcolor{green!50} \textbf{0.6} 
 \\ \hline
cycle 
& \textbf{0.8} 
& \textbf{0.8} 
& 0.05 
& \cellcolor{yellow!50} \textbf{0.8} 
& \cellcolor{yellow!50} \textbf{0.8} 
& \cellcolor{green!50} \textbf{0.8} 
 \\ \hline
tree 
& 0.61 
& 0.56 
& 0.04 
& \cellcolor{green!50} 0.78 
& \cellcolor{green!50} 0.83 
& \cellcolor{green!50} \textbf{0.88} 
 \\ \hline
block 
& 0.05 
& 0.01 
& 0.0 
& \cellcolor{green!50} \textbf{0.36} 
& \cellcolor{green!50} 0.02 
& \cellcolor{green!50} 0.29 
 \\ \hline
compl. 
& 0.0 
& \textbf{0.01} 
& 0.0 
& \cellcolor{yellow!50} 0.0 
& \cellcolor{yellow!50} \textbf{0.01} 
& \cellcolor{yellow!50} 0.0 
 \\ \hline
cube 
& 0.28 
& 0.3 
& 0.01 
& \cellcolor{green!50} \textbf{0.46} 
& \cellcolor{green!50} 0.44 
& \cellcolor{green!50} 0.4 
 \\ \hline
symme. 
& 0.66 
& 0.6 
& 0.03 
& \cellcolor{green!50} 0.68 
& \cellcolor{green!50} 0.76 
& \cellcolor{green!50} \textbf{0.77} 
 \\ \hline
bipar. 
& 0.01 
& 0.03 
& 0.01 
& \cellcolor{green!50} 0.02 
& \cellcolor{green!50} 0.04 
& \cellcolor{green!50} \textbf{0.11} 
 \\ \hline
grid 
& 0.52 
& \textbf{0.54} 
& 0.0 
& \cellcolor{yellow!50} 0.52 
& \cellcolor{yellow!50} \textbf{0.54} 
& \cellcolor{green!50} 0.52 
 \\ \hline
spx t. 
& 0.02 
& 0.0 
& 0.0 
& \cellcolor{green!50} \textbf{0.03} 
& \cellcolor{yellow!50} 0.0 
& \cellcolor{yellow!50} 0.0 
 \\ \hline
\end{tabular}
}
\end{minipage}
\par\medskip
\label{tab:avg_angular_resolution}
\begin{minipage}{.5\linewidth}
\centering
{\tiny
\begin{tabular}{|l|l|l|l|l|l|l|} \hline
\multicolumn{7}{|c|}{
Neighbor preservation
}\\ \hline
\multicolumn{1}{|c|}{} & \multicolumn{1}{c|}{neato} & \multicolumn{1}{c|}{sdfp} & rnd & \multicolumn{1}{c|}{$(GD)^2_n$} & \multicolumn{1}{c|}{$(GD)^2_s$} & \multicolumn{1}{c|}{$(GD)^2_r$} \\ \hline
dodec. 
& 0.32 
& 0.3 
& 0.1 
& \cellcolor{green!50} \textbf{0.5} 
& \cellcolor{yellow!50} 0.3 
& \cellcolor{green!50} \textbf{0.5} 
 \\ \hline
cycle 
& \textbf{1.0} 
& \textbf{1.0} 
& 0.08 
& \cellcolor{yellow!50} \textbf{1.0} 
& \cellcolor{yellow!50} \textbf{1.0} 
& \cellcolor{green!50} \textbf{1.0} 
 \\ \hline
tree 
& \textbf{1.0} 
& \textbf{1.0} 
& 0.02 
& \cellcolor{yellow!50} \textbf{1.0} 
& \cellcolor{yellow!50} \textbf{1.0} 
& \cellcolor{green!50} \textbf{1.0} 
 \\ \hline
block 
& 0.57 
& 0.93 
& 0.12 
& \cellcolor{green!50} 0.83 
& \cellcolor{yellow!50} 0.93 
& \cellcolor{green!50} \textbf{1.0} 
 \\ \hline
compl. 
& \textbf{1.0} 
& \textbf{1.0} 
& \textbf{1.0} 
& \cellcolor{yellow!50} \textbf{1.0} 
& \cellcolor{yellow!50} \textbf{1.0} 
& \cellcolor{yellow!50} \textbf{1.0} 
 \\ \hline
cube 
& \textbf{0.5} 
& \textbf{0.5} 
& 0.12 
& \cellcolor{yellow!50} \textbf{0.5} 
& \cellcolor{yellow!50} \textbf{0.5} 
& \cellcolor{green!50} \textbf{0.5} 
 \\ \hline
symme. 
& 0.75 
& 0.95 
& 0.05 
& \cellcolor{yellow!50} 0.75 
& \cellcolor{green!50} \textbf{1.0} 
& \cellcolor{green!50} \textbf{1.0} 
 \\ \hline
bipar. 
& \textbf{0.47} 
& \textbf{0.47} 
& 0.43 
& \cellcolor{yellow!50} \textbf{0.47} 
& \cellcolor{yellow!50} \textbf{0.47} 
& \cellcolor{yellow!50} 0.43 
 \\ \hline
grid 
& \textbf{1.0} 
& \textbf{1.0} 
& 0.05 
& \cellcolor{yellow!50} \textbf{1.0} 
& \cellcolor{yellow!50} \textbf{1.0} 
& \cellcolor{green!50} \textbf{1.0} 
 \\ \hline
spx t. 
& 0.36 
& 0.44 
& 0.03 
& \cellcolor{green!50} 0.49 
& \cellcolor{green!50} 0.46 
& \cellcolor{green!50} \textbf{0.53} 
 \\ \hline
\end{tabular}
}
\end{minipage}
\label{tab:avg_neighbor}
\begin{minipage}{.5\linewidth}
\centering
{\tiny
\begin{tabular}{|l|l|l|l|l|l|l|} \hline
\multicolumn{7}{|c|}{
Gabriel graph property
}\\ \hline
\multicolumn{1}{|c|}{} & \multicolumn{1}{c|}{neato} & \multicolumn{1}{c|}{sdfp} & rnd & \multicolumn{1}{c|}{$(GD)^2_n$} & \multicolumn{1}{c|}{$(GD)^2_s$} & \multicolumn{1}{c|}{$(GD)^2_r$} \\ \hline
dodec. 
& 0.16 
& \textbf{0.64} 
& 0.07 
& \cellcolor{green!50} 0.32 
& \cellcolor{yellow!50} \textbf{0.64} 
& \cellcolor{green!50} 0.32 
 \\ \hline
cycle 
& \textbf{1.0} 
& \textbf{1.0} 
& 0.29 
& \cellcolor{yellow!50} \textbf{1.0} 
& \cellcolor{yellow!50} \textbf{1.0} 
& \cellcolor{green!50} \textbf{1.0} 
 \\ \hline
tree 
& \textbf{1.0} 
& \textbf{1.0} 
& 0.05 
& \cellcolor{yellow!50} \textbf{1.0} 
& \cellcolor{yellow!50} \textbf{1.0} 
& \cellcolor{green!50} \textbf{1.0} 
 \\ \hline
block 
& 0.16 
& 0.03 
& 0.04 
& \cellcolor{green!50} 0.57 
& \cellcolor{green!50} 0.14 
& \cellcolor{green!50} \textbf{0.59} 
 \\ \hline
compl. 
& 0.0 
& 0.01 
& 0.02 
& \cellcolor{green!50} 0.04 
& \cellcolor{yellow!50} 0.01 
& \cellcolor{green!50} \textbf{0.07} 
 \\ \hline
cube 
& 0.43 
& 0.51 
& 0.01 
& \cellcolor{green!50} 0.75 
& \cellcolor{green!50} \textbf{0.8} 
& \cellcolor{green!50} 0.71 
 \\ \hline
symme. 
& 0.54 
& \textbf{1.0} 
& 0.15 
& \cellcolor{green!50} 0.7 
& \cellcolor{yellow!50} \textbf{1.0} 
& \cellcolor{green!50} \textbf{1.0} 
 \\ \hline
bipar. 
& 0.08 
& 0.11 
& 0.25 
& \cellcolor{green!50} 0.48 
& \cellcolor{green!50} 0.64 
& \cellcolor{green!50} \textbf{0.74} 
 \\ \hline
grid 
& \textbf{1.0} 
& \textbf{1.0} 
& 0.03 
& \cellcolor{yellow!50} \textbf{1.0} 
& \cellcolor{yellow!50} \textbf{1.0} 
& \cellcolor{green!50} \textbf{1.0} 
 \\ \hline
spx t. 
& 0.04 
& 0.0 
& 0.02 
& \cellcolor{green!50} 0.06 
& \cellcolor{green!50} \textbf{0.08} 
& \cellcolor{green!50} \textbf{0.08} 
 \\ \hline
\end{tabular}
}
\end{minipage}
\par\medskip
\label{tab:avg_gabriel}
\begin{minipage}{.5\linewidth}
\centering
{\tiny
\begin{tabular}{|l|l|l|l|l|l|l|} \hline
\multicolumn{7}{|c|}{
Vertex resolution
}\\ \hline
\multicolumn{1}{|c|}{} & \multicolumn{1}{c|}{neato} & \multicolumn{1}{c|}{sdfp} & rnd & \multicolumn{1}{c|}{$(GD)^2_n$} & \multicolumn{1}{c|}{$(GD)^2_s$} & \multicolumn{1}{c|}{$(GD)^2_r$} \\ \hline
dodec. 
& 0.52 
& 0.54 
& 0.07 
& \cellcolor{green!50} 0.7 
& \cellcolor{green!50} \textbf{0.81} 
& \cellcolor{green!50} 0.68 
 \\ \hline
cycle 
& \textbf{0.98} 
& \textbf{0.98} 
& 0.32 
& \cellcolor{yellow!50} \textbf{0.98} 
& \cellcolor{yellow!50} \textbf{0.98} 
& \cellcolor{green!50} \textbf{0.98} 
 \\ \hline
tree 
& 0.68 
& 0.57 
& 0.23 
& \cellcolor{green!50} \textbf{0.69} 
& \cellcolor{green!50} 0.68 
& \cellcolor{green!50} 0.68 
 \\ \hline
block 
& 0.66 
& 0.38 
& 0.1 
& \cellcolor{green!50} \textbf{0.72} 
& \cellcolor{green!50} 0.59 
& \cellcolor{green!50} 0.51 
 \\ \hline
compl. 
& 0.8 
& \textbf{1.0} 
& 0.18 
& \cellcolor{green!50} 0.84 
& \cellcolor{yellow!50} \textbf{1.0} 
& \cellcolor{green!50} 0.91 
 \\ \hline
cube 
& 0.66 
& \textbf{0.82} 
& 0.11 
& \cellcolor{yellow!50} 0.66 
& \cellcolor{yellow!50} \textbf{0.82} 
& \cellcolor{green!50} 0.67 
 \\ \hline
symme. 
& 0.35 
& 0.43 
& 0.06 
& \cellcolor{green!50} 0.38 
& \cellcolor{green!50} 0.51 
& \cellcolor{green!50} \textbf{0.6} 
 \\ \hline
bipar. 
& 0.83 
& \textbf{0.87} 
& 0.21 
& \cellcolor{yellow!50} 0.83 
& \cellcolor{yellow!50} \textbf{0.87} 
& \cellcolor{green!50} 0.35 
 \\ \hline
grid 
& 0.87 
& 0.8 
& 0.08 
& \cellcolor{green!50} \textbf{0.88} 
& \cellcolor{green!50} \textbf{0.88} 
& \cellcolor{green!50} \textbf{0.88} 
 \\ \hline
spx t. 
& 0.47 
& \textbf{0.48} 
& 0.05 
& \cellcolor{yellow!50} 0.47 
& \cellcolor{yellow!50} \textbf{0.48} 
& \cellcolor{green!50} 0.32 
 \\ \hline
\end{tabular}
}
\end{minipage}
\label{tab:avg_vertex_resolution}
\begin{minipage}{.5\linewidth}
\centering
{\tiny
\begin{tabular}{|l|l|l|l|l|l|l|} \hline
\multicolumn{7}{|c|}{
Aspect ratio
}\\ \hline
\multicolumn{1}{|c|}{} & \multicolumn{1}{c|}{neato} & \multicolumn{1}{c|}{sdfp} & rnd & \multicolumn{1}{c|}{$(GD)^2_n$} & \multicolumn{1}{c|}{$(GD)^2_s$} & \multicolumn{1}{c|}{$(GD)^2_r$} \\ \hline
dodec. 
& 0.92 
& 0.91 
& 0.88 
& \cellcolor{green!50} \textbf{0.96} 
& \cellcolor{green!50} \textbf{0.96} 
& \cellcolor{green!50} \textbf{0.96} 
 \\ \hline
cycle 
& \textbf{0.96} 
& 0.95 
& 0.67 
& \cellcolor{yellow!50} \textbf{0.96} 
& \cellcolor{yellow!50} 0.95 
& \cellcolor{green!50} \textbf{0.96} 
 \\ \hline
tree 
& 0.73 
& 0.67 
& \textbf{0.88} 
& \cellcolor{green!50} 0.86 
& \cellcolor{green!50} 0.76 
& \cellcolor{yellow!50} \textbf{0.88} 
 \\ \hline
block 
& 0.9 
& 0.74 
& 0.7 
& \cellcolor{green!50} \textbf{0.96} 
& \cellcolor{green!50} 0.9 
& \cellcolor{green!50} \textbf{0.96} 
 \\ \hline
compl. 
& 0.89 
& 0.97 
& 0.91 
& \cellcolor{green!50} \textbf{0.98} 
& \cellcolor{green!50} \textbf{0.98} 
& \cellcolor{green!50} \textbf{0.98} 
 \\ \hline
cube 
& 0.76 
& 0.79 
& 0.57 
& \cellcolor{green!50} 0.87 
& \cellcolor{yellow!50} 0.79 
& \cellcolor{green!50} \textbf{0.88} 
 \\ \hline
symme. 
& 0.58 
& 0.67 
& \textbf{0.89} 
& \cellcolor{green!50} 0.6 
& \cellcolor{yellow!50} 0.67 
& \cellcolor{yellow!50} \textbf{0.89} 
 \\ \hline
bipar. 
& 0.82 
& 0.9 
& \textbf{0.91} 
& \cellcolor{yellow!50} 0.82 
& \cellcolor{yellow!50} 0.9 
& \cellcolor{yellow!50} \textbf{0.91} 
 \\ \hline
grid 
& \textbf{1.0} 
& \textbf{1.0} 
& 0.82 
& \cellcolor{yellow!50} \textbf{1.0} 
& \cellcolor{yellow!50} \textbf{1.0} 
& \cellcolor{green!50} \textbf{1.0} 
 \\ \hline
spx t. 
& 0.98 
& 0.86 
& 0.88 
& \cellcolor{green!50} \textbf{0.99} 
& \cellcolor{green!50} \textbf{0.99} 
& \cellcolor{green!50} \textbf{0.99} 
 \\ \hline
\end{tabular}
}
\end{minipage}
\par\medskip
\label{tab:avg_aspect_ratio}
\centering
\begin{minipage}{.5\linewidth}
{\tiny
\begin{tabular}{|l|l|l|l|l|l|l|} \hline
\multicolumn{7}{|c|}{
Crossing angle
}\\ \hline
\multicolumn{1}{|c|}{} & \multicolumn{1}{c|}{neato} & \multicolumn{1}{c|}{sdfp} & rnd & \multicolumn{1}{c|}{$(GD)^2_n$} & \multicolumn{1}{c|}{$(GD)^2_s$} & \multicolumn{1}{c|}{$(GD)^2_r$} \\ \hline
dodec. 
& \textbf{0.06} 
& 0.12 
& 0.24 
& \cellcolor{yellow!50} \textbf{0.06} 
& \cellcolor{green!50} 0.09 
& \cellcolor{green!50} 0.15 
 \\ \hline
cycle 
& \textbf{0.0} 
& \textbf{0.0} 
& 0.19 
& \cellcolor{yellow!50} \textbf{0.0} 
& \cellcolor{yellow!50} \textbf{0.0} 
& \cellcolor{green!50} \textbf{0.0} 
 \\ \hline
tree 
& \textbf{0.0} 
& \textbf{0.0} 
& 0.23 
& \cellcolor{yellow!50} \textbf{0.0} 
& \cellcolor{yellow!50} \textbf{0.0} 
& \cellcolor{green!50} \textbf{0.0} 
 \\ \hline
block 
& 0.11 
& 0.1 
& 0.24 
& \cellcolor{green!50} \textbf{0.05} 
& \cellcolor{green!50} 0.06 
& \cellcolor{green!50} 0.09 
 \\ \hline
compl. 
& 0.25 
& \textbf{0.24} 
& \textbf{0.24} 
& \cellcolor{green!50} \textbf{0.24} 
& \cellcolor{yellow!50} \textbf{0.24} 
& \cellcolor{yellow!50} \textbf{0.24} 
 \\ \hline
cube 
& \textbf{0.03} 
& \textbf{0.03} 
& 0.21 
& \cellcolor{yellow!50} \textbf{0.03} 
& \cellcolor{yellow!50} \textbf{0.03} 
& \cellcolor{green!50} 0.04 
 \\ \hline
symme. 
& 0.03 
& \textbf{0.0} 
& 0.24 
& \cellcolor{yellow!50} 0.03 
& \cellcolor{yellow!50} \textbf{0.0} 
& \cellcolor{green!50} \textbf{0.0} 
 \\ \hline
bipar. 
& \textbf{0.16} 
& 0.17 
& 0.23 
& \cellcolor{yellow!50} \textbf{0.16} 
& \cellcolor{yellow!50} 0.17 
& \cellcolor{green!50} 0.19 
 \\ \hline
grid 
& \textbf{0.0} 
& \textbf{0.0} 
& 0.23 
& \cellcolor{yellow!50} \textbf{0.0} 
& \cellcolor{yellow!50} \textbf{0.0} 
& \cellcolor{green!50} \textbf{0.0} 
 \\ \hline
spx t. 
& 0.16 
& 0.22 
& 0.25 
& \cellcolor{yellow!50} 0.16 
& \cellcolor{green!50} \textbf{0.15} 
& \cellcolor{green!50} 0.21 
 \\ \hline
\end{tabular}
}
\end{minipage}
\caption{The values of the nine criteria corresponding to the 10 graphs for the layouts computed by neato, sfdp, random, and 3 runs of \GDGD initialized with neato, sfdp, and random layouts. Bold values are the best. Green cells show an improvement, yellow cells show a tie,
 with respect to the initial values.
 }
\label{tab:avg_crossing_angle}
\end{table}
\bibliography{references}
\bibliographystyle{splncs04}

\newpage\section{Appendix}
The following table summarizes the  objective functions used to optimize the nine drawing criteria via different optimization methods.
\begin{table}[]
\resizebox{1.1\textwidth}{!}{
\centering
\begin{tabular}{p{2.0cm}|p{4cm}|p{4cm}|p{4cm}}
\toprule
Property 
& Gradient Descent 
& Subgradient Descent 
& Stochastic Gradient Descent\\
\midrule
Stress
 & $\sum\limits_{i<j}\;w_{ij}(|X_i - X_j|_2 - d_{ij})^2$ 
& $\sum\limits_{i<j}\;w_{ij}(|X_i - X_j|_2 - d_{ij})^2$ 
& $w_{ij}(|X_i - X_j|_2 - d_{ij})^2$ for a random pair of nodes $i, j \in V$\\ 
\hline

Ideal \hspace{.8cm}Edge Length
& $\sqrt{ \frac{1}{|E|} \sum\limits_{(i,j) \in E}\;  
(\frac{||X_i - X_j|| - l_{ij}}{l_{ij}}})^2$ (Eq.~\ref{eq:loss-ideal-edge-length})
& $\frac{1}{|E|} \sum\limits_{(i,j) \in E}\;  
|\frac{||X_i - X_j|| - l_{ij}}{l_{ij}}|$
& $|\frac{||X_i - X_j|| - l_{ij}}{l_{ij}}|$ for a random edge $(i,j) \in E$\\
\hline

Crossing \hspace{.5cm} Angle 
& $\sum\limits_i\;cos(\theta_i)^2$ 
& $\sum\limits_i\; |cos(\theta_i)|$
& $|cos(\theta_i)|$ for a random crossing $i$\\
\hline

Neighborhood Preservation
& Lov\'asz \textbf{softmax}~\cite{berman2018lovasz} between
  neighborhood prediction (Eq.\ref{eq:neighbor-pred})
  and adjacency matrix $Adj$
& Lov\'asz \textbf{hinge}~\cite{berman2018lovasz} between 
  neighborhood prediction (Eq.\ref{eq:neighbor-pred}) 
  and adjacency matrix $Adj$
& Lov\'asz \textbf{softmax} or \textbf{hinge}~\cite{berman2018lovasz} on a random node. 
  (i.e. Jaccard loss between a random \textit{row} of K in Eq. \ref{eq:neighbor-pred} 
  and the corresponding row in the adjacency matrix $Adj$)\\
\hline

Crossing Number
& Shabbeer et al.~\cite{bennett2010}
& Shabbeer et al.~\cite{bennett2010}
& Shabbeer et al.~\cite{bennett2010}\\
\hline

Angular \hspace{.5cm} Resolution
& $\sum\limits_{(i,j),(j,k) \in E}\; e^{-\varphi_{ijk}}$
& $\sum\limits_{v \in E}\; e^{-\varphi_{ijk}}$
& \makecell[l]{$e^{-\varphi_{ijk}}$ \\for random $(i,j),(j,k) \in E$} \\
\hline

Vertex\hspace{.8cm} Resolution
& $\sum_{i,j \in V, i \neq j}$ ${ReLU( 1 - \frac{||X_i - X_j||}{d_{max} \cdot r}) ^2}$ (Eq. \ref{eq:loss-vertex-resolution})
& $\sum_{i,j \in V, i \neq j}$ ${ReLU( 1 - \frac{||X_i - X_j||}{d_{max} \cdot r})}$
& $ReLU( 1 - \frac{||X_i - X_j||}{d_{max} \cdot r})$ for random $i,j \in V, i \neq j$
\\
\hline

Gabriel Graph
&$\sum_{\substack{(i,j) \in E, k \in V \setminus \{i,j\}}} $ $ReLU(r_{ij} - |X_k - c_{ij}|) \; ^ 2$ (Eq. \ref{eq:gabriel})
&$\sum_{\substack{(i,j) \in E,k \in V \setminus \{i,j\}}} $ $ReLU(r_{ij} - |X_k - c_{ij}|)$
&$ReLU(r_{ij} - |X_k - c_{ij}|)$ for random $(i,j) \in E$ and $k \in V \setminus \{i,j\}$ 
\\
\hline

Aspect Ratio
& Eq. \ref{eq:aspect-ratio}
& Eq. \ref{eq:aspect-ratio}
& Eq. \ref{eq:aspect-ratio}
\\
\bottomrule\\
\end{tabular}
} 
\caption{Summary of the objective functions via different optimization methods.}
\label{table:loss-functions}
\end{table}

\end{document}